\documentclass[12pt]{l4dc2020} 

\title[Dual Stochastic MPC]{Dual Stochastic MPC for Systems with \\ Parametric and Structural Uncertainty }
\usepackage{times}
\usepackage{amsmath}
\usepackage{amssymb}

\usepackage{tikz}
\usetikzlibrary{decorations.pathreplacing}
\usepackage{pgfplots}
\usepackage{xcolor}
\usepackage[short]{optidef}

\newcommand{\version}[2]{#1} 

\author{%
 \Name{Elena Arcari}\textsuperscript{1} \Email{earcari@ethz.ch}\\ 
 \Name{Lukas Hewing}\textsuperscript{1} \Email{lhewing@ethz.ch}  \\
 \Name{Max Schlichting}\textsuperscript{2} \Email{maxschlichting@web.de}\\
 \Name{Melanie N. Zeilinger}\textsuperscript{1} \Email{mzeilinger@ethz.ch}\\
 \addr \textsuperscript{1} Institute for Dynamic Systems and Control, ETH Zurich, Zurich, Switzerland \\
 \textsuperscript{2} Department of Neurorobotics, University of Freiburg, Freiburg, Germany
 }
 
\begin{document}

\maketitle

\begin{abstract}%
  Designing controllers for systems affected by model uncertainty can prove to be a challenge, especially when seeking the optimal compromise between the conflicting goals of identification and control. This trade-off is explicitly taken into account in the dual control problem, for which the exact solution is provided by stochastic dynamic programming. Due to its computational intractability, we propose a sampling-based approximation for systems affected by both parametric and structural model uncertainty. The approach proposed in this paper separates the prediction horizon in a dual and an exploitation part. The dual part is formulated as a scenario tree that actively discriminates among a set of potential models while learning unknown parameters. In the exploitation part,  achieved information is fixed for each scenario, and open-loop control sequences are computed for the remainder of the horizon. As a result, we solve one optimization problem over a collection of control sequences for the entire horizon, explicitly considering the knowledge gained in each scenario, leading to a dual model predictive control formulation.  
\end{abstract}


\section{Introduction}

Model-based control techniques are commonly used to enable complex systems to perform challenging tasks, however, their performance strongly depends on the model accuracy. Adaptive control methods address the problem of simultaneous learning and control, avoiding the need for costly offline (re-)identification. Classic adaptive control is typically based on a single model with \emph{parametric} uncertainty~\citep{KJA94,SS90}. This description can be insufficient when the effect of the uncertainty is large, and modifies the overall structure of the model. For such \emph{structural} uncertainty~\citep{RM97,KSN11}, it is beneficial to highlight several operating modes that, for instance,  distinguish nominal behavior from fault conditions. In multiple-model approaches, a controller is designed to switch online between a finite number of models representing distinct system behaviors in order to ensure adequate control under each possible scenario. Available techniques, however, typically estimate the current operating mode of the system, and switch between pre-defined model-based controllers, but do not enable \emph{active} learning about structural uncertainty, or consider the trade-off between several likely models. For a comprehensive review of multiple-model control we refer to ~\citep{TANH19}. 

\emph{Active} adaptive control schemes explicitly take into account the probing effect that the control input has on the system, and therefore the knowledge to be gained over structural or parametric uncertainty. Excitation needed to learn the model is optimally traded-off with the control performance when applying dual control~\citep{AAF60}, for which the exact solution is provided by stochastic dynamic programming (DP)~\citep{DB17}. Due to its computational complexity, developed techniques revert to approximations, which can be classified into two categories: explicit and implicit approaches. Explicit techniques~\citep{GM14,TANH17} generally introduce a heuristic term in the cost function that drives the input to explore. These methods are usually computationally efficient, but require extensive expert knowledge and explore heuristically, e.g. often the entire system is excited for identification regardless of the benefit for the specific task. Implicit techniques~\citep{ET73,DB85,EK16}, on the other hand, are based on analytic reformulations that give a theoretically sound approximation of the original dual control problem, but are usually computationally expensive. Since implicit methods are developed in a generic framework, they can in principle be applied to a large class of systems, while ensuring the dual effect as an intrinsic property. We refer to~\citep{AM17} for a more detailed overview of dual control.

In this paper, we develop an implicit dual control scheme, by using a sampling method based on approximate dynamic programming (ADP), and making use of a rollout approach~\citep{DB17}. Samples are used to generate a scenario tree associated with an optimization problem that provides optimal control sequences for each scenario, taking the gained knowledge explicitly into account. This formulation gives rise to a dual stochastic model predictive control problem that builds on ideas included in~\citep{EA19} and~\citep{KGH15}, in which similar schemes are applied to systems linear in the uncertain parameters. In this paper, the formulation is extended to the important case of systems with uncertainty in both the parameters and in the model structure. The performance of the approach is shown in a simulation example of an aircraft and compared to model predictive control (MPC) with passive adaptation.

\section{Preliminaries}

\subsection{Problem Formulation}
\label{Problem Formulation}

We consider nonlinear dynamic systems, which are characterized by a finite set of operating modes $\mathcal{M} = \{ M^1, \dots, M^{n_m}\}$, e.g.\ corresponding to different failure cases. Each operating mode $M \in \mathcal{M}$ is associated with a prior probability $p(M)$ of being active, and the system dynamics given mode $M$ are
\begin{equation}
M : \  x_{k+1} = \Phi^{M} (x_k,u_k)\gamma^{M} + w^{M}_k,
\label{eq:model}
\end{equation}
with state $x_k \in \mathbb{R}^{n_x}$ and control input $u_k \in \mathcal{U} \subseteq
\mathbb{R}^{n_u}$, where $\mathcal{U}$ defines the input constraints. Each mode is described
by a nonlinear basis function matrix $\Phi^{M}: \mathbb{R}^{n_x} \times \mathbb{R}^{n_u} \rightarrow
\mathbb{R}^{n_x \times n_\gamma^{M}}$ with associated uncertain parameters $\gamma^{M} \in
\mathbb{R}^{n_\gamma^{M}}$, which we assume to be distributed
according to $\gamma^{M} \sim \mathcal{N}(\mu_\gamma^{M}, \Sigma_\gamma^{M})$. The system is
furthermore subject to additive disturbances $w^{M}_k \sim \mathcal{N}(0,\Sigma^{M}_w)$, which we
consider zero mean i.i.d.\ for (notational) simplicity. At each time step $k$, we assume access to
measurements of the state $x_k$, which are used to identify both the uncertain operating mode $M$
and the corresponding parameterization $\gamma^{M}$.
\begin{remark}
  The structural uncertainty description through a collection of modes $\mathcal{M}$ is applicable to a wide range of use cases. This includes, for instance, systems affected by parameters taking discrete values, multimodal parameter distributions or discretizations of continuous uncertain parameters which do not allow for an expression in the form of $\gamma$ in \eqref{eq:model}.
\end{remark}
The considered objective is to find an optimal policy that simultaneously controls the system and identifies its operating mode and corresponding unknown parameters. For this purpose, we define a finite-horizon cost
\begin{equation}
  J_{\bar{N}}(\Pi,x_0) := \underset{\substack{M, \gamma^M, \\ w^M_0,\ldots, w^M_{\bar{N}-1}}}{\mathbb{E}} \sum_{k=0}^{\bar{N}} l_k(x_k,\pi_k(x_k)),
  \label{eq:cost_function}
  \end{equation} 
where $\Pi = \{ \pi_0(\cdot), \dots, \pi_{\bar{N}-1}(\cdot) \}$ is a policy sequence that we want to optimize by minimizing~\eqref{eq:cost_function}, $\bar{N}$ is the length of the overall control task and $l_k: \mathbb{R}^{n_x} \times \mathbb{R}^{n_u} \rightarrow \mathbb{R}$ is a potentially time-varying cost function. As a result, we obtain a finite-horizon stochastic optimal control problem that can be addressed using dynamic programming (DP), which offers the property of generating dual control policies. In the following section we provide a DP formulation that handles both parametric and structural uncertainty, and subsequently outline a tractable approximate DP scheme.

\subsection{Stochastic DP with Parametric and Structural Model Uncertainty}
\label{Stochastic DP with Structural Model Uncertainty}

In order to address the problem by DP, we provide a formulation in the context of systems with imperfect state information~\citep{DB17}. We define the information vector $\mathcal{I}_k$, expressing the available information at time step $k$ recursively given a known initial state , i.e.\ $\mathcal{I}_0 = x_0$, as
\begin{equation*}
\mathcal{I}_{k+1} = \left[ x_{k+1}^T, u_{k}^T, \mathcal{I}_{k}^T \right]^T.
\end{equation*}
The information at time step $k$ takes the function of the state in the DP recursion
\begin{equation}
J_k^*(\mathcal{I}_{k}) = \displaystyle\min_{\pi_k} \ l_k(x_k,\pi_k) + \underset{M,\gamma^M,w^M_k}{\mathbb{E}} \left[ J_{k+1}^*(\mathcal{I}_{k+1}) \ | \ \mathcal{I}_k \right ], \quad k = 0,\dots, \bar{N}-1
\label{eq:DP_recursion}
\end{equation}
initialized at $J_{\bar{N}}^*(\mathcal{I}_{\bar{N}}) = l_{\bar{N}}(x_{\bar{N}})$. The expected value arising in the recursion is computed with respect to the currently available information $\mathcal{I}_k$, which means that the parameter distribution is accordingly updated at each time step. When receiving a new state measurement $x_{k+1}$, which updates the current information vector to $\mathcal{I}_{k+1}$, the posterior joint distribution of mode $M$ and $\gamma^{M}$ is given by:
  \begin{equation*}
  p(\gamma^{M}, M \, | \, \mathcal{I}_{k+1}) = p(\gamma^{M} \, | \, \mathcal{I}_{k+1}, M)p(M \, | \, \mathcal{I}_{k+1}),
  \end{equation*}  
where $p(M \, | \, \mathcal{I}_{k+1})$ is the probability that mode $M$ is active given the information at time step $k+1$, and $p(\gamma^{M}| \mathcal{I}_{k+1}, M)$ is the distribution of the associated parameters. Each component is updated via Bayesian estimation~\citep{AM17}:
\begin{subequations} \label{eq:Bayesian_update}
\begin{align}
p(\gamma^{M}\,|\, \mathcal{I}_{k+1},M) &= \frac{p(x_{k+1} \,|\, u_{k}, \mathcal{I}_k, M, \gamma^{M}) p(\gamma^{M}\,|\,  \mathcal{I}_{k}, M)}{p(x_{k+1} \,|\, u_k, \mathcal{I}_k, M)}, \label{eq:Bayesian_update_param}
\\
p(M \,|\, \mathcal{I}_{k+1}) & = \frac{p(x_{k+1} \,|\, u_{k}, \mathcal{I}_{k}, M) p(M \,|\, \mathcal{I}_{k})}{p(x_{k+1} \, | \, u_{k} ,\mathcal{I}_{k})}.
\label{eq:Bayesian_update_model}
\end{align}
\end{subequations}
The parameter update given mode $M$ in~\eqref{eq:Bayesian_update_param} follows Gaussian linear
regression, in which the likelihood $p(x_{k+1} \,|\, u_{k}, \mathcal{I}_k, M, \gamma^{M})$ can be computed from the corresponding Gaussian noise distribution in~\eqref{eq:model}. The normalization factor $p(x_{k+1} \,|\, u_k, \mathcal{I}_k, M)$ is subsequently used to evaluate the likelihood of each individual mode in~\eqref{eq:Bayesian_update_model}. For details on the Bayesian estimation procedure we refer to \version{Appendix~\ref{app:bayesian estimation}}{Appendix A in~\cite{EA19b}}.
\begin{remark}
 For long run times $\bar{N}$, the formulation above leads to convergence of both the operating mode and parameter estimate to absolute certainty, due to the accumulated information. This results in a loss of adaptability, which can be undesired e.g.\ for fault detection. Convergence can be avoided, for instance, by introducing process noise on the parameter $\gamma^M$ and switching between modes according to a Markov process, resulting in increased complexity of the estimation in~\eqref{eq:Bayesian_update}, see e.g.\ \citep{PSM79}. As a computationally efficient alternative, we heuristically cap the probability of each mode, as well as the variance of the parameter estimate.
 \label{capping remark}
  \end{remark}
While DP provides the exact solution to the dual control problem in the case of both parametric and structural model uncertainty, it is generally tractable only for very small problems. We provide a tractable approximate approach based on~\citep{EA19}, which we extend to the case of both continuous parameters and structural uncertainty.

\section{Dual Stochastic MPC}

The proposed formulation is based on a receding horizon implementation of an approximate dynamic programming (ADP) strategy. We repeatedly solve an approximation of problem~\eqref{eq:cost_function} over a shortened horizon $N \ll \bar{N}$, and apply the first computed control input, i.e. using receding horizon control. To approximate the receding horizon problem, we furthermore use a rollout approach~\citep{DB17}, and truncate the prediction horizon of length $N$ to $L<N$. The control sequence associated with the truncated part of the horizon of length $N-L$ is optimized in open-loop, as in model predictive control, and defines a suboptimal cost-to-go $\tilde J_L(\mathcal{I}_L)$. The first part of the horizon of length $L$, is solved by approximating the DP recursion~\eqref{eq:DP_recursion} using a sampling-based approach. The expected values arising at each DP iteration in~\eqref{eq:DP_recursion} are generally not available in closed form, even if the Bayesian updates in~\eqref{eq:Bayesian_update} can be evaluated exactly, e.g.\ with self-conjugate distributions~\citep{EK16}. For this reason, in~\cite{EA19}, expected values are evaluated as averages over parameter and noise samples. We use a similar idea for handling the case of both continuous parameters and structural model uncertainty, such that it preserves the dual properties of the DP policy. In the following subsections, we outline the sampling approach used for approximating DP along the first $L$ steps of the prediction horizon, which we refer to as \emph{dual} part, and we discuss the cost-to-go $\tilde J_L(\mathcal{I}_L)$ that is solved in the \emph{exploitation} part, covering the remaining part of the horizon of length $N-L$.

\subsection{Dual Part}
\label{Dual Part}

\begin{figure}[t]
\centering
\begin{tikzpicture}[
c0/.style n args={2}{insert path={node[n0={#1}{#2}] (n0#1#2){}}},
n0/.style n args={2}{circle,fill,inner sep=1pt,label={90:\footnotesize $x_{#1}^{#2}$}},
cmid/.style 2 args={insert path={node[nmid={#1}{#2}] (nmid#1#2){}}},
nmid/.style={circle,fill,inner sep=0pt}, scale={0.5},
]
\path (0,0)[c0={0}{1}];
\draw (n001) -- node[above]  {\footnotesize $M^1$} (2,5)[cmid={1}{1}];
\draw (n001) -- node[below] {\footnotesize $M^2$}(2,-5)[cmid={1}{2}];
\draw (nmid11) -- node[above] {\footnotesize $w_0^1, \gamma_0^1$}(6,7)[c0={1}{1}];
\draw (nmid11) -- node[above] {\footnotesize $w_0^2, \gamma_0^2$}(6,4)[c0={1}{2}];
\draw (nmid12) -- node[above] {\footnotesize $w_0^3, \gamma_0^3$}(6,-4)[c0={1}{3}];
\draw (nmid12) -- node[above] {\footnotesize $w_0^4, \gamma_0^4$}(6,-7)[c0={1}{4}];
\draw (n012) -- node[above] {\footnotesize $M^1$} (9,6) [cmid={2}{1}];
\draw (n012) -- node[above] {\footnotesize $M^2$} (9,2) [cmid={2}{2}];
\draw (n013) -- node[above] {\footnotesize $M^1$} (9,-2) [cmid={3}{1}];
\draw (n013) -- node[above] {\footnotesize $M^2$} (9,-6) [cmid={3}{2}];
\draw (nmid21) -- node[above] {\footnotesize $w_1^5, \gamma_1^5$} (13,7) [c0={2}{5}];
\draw (nmid21) -- node[below] {\footnotesize $w_1^6, \gamma_1^6$} (13,5) [c0={2}{6}];
\draw (nmid22) -- node[above] {\footnotesize $w_1^7, \gamma_1^7$} (13,3) [c0={2}{7}];
\draw (nmid22) -- node[below] {\footnotesize $w_1^8, \gamma_1^8$} (13,1) [c0={2}{8}];
\draw (nmid31) -- node[above] {\footnotesize $w_1^9, \gamma_1^9$} (13,-1) [c0={2}{9}];
\draw (nmid31) -- node[below] {\footnotesize $w_1^{10}, \gamma_1^{10}$} (13,-3) [c0={2}{10}];
\draw (nmid32) -- node[above] {\footnotesize $w_1^{11}, \gamma_1^{11}$} (13,-5) [c0={2}{11}];
\draw (nmid32) -- node[below] {\footnotesize $w_1^{12}, \gamma_1^{12}$} (13,-7) [c0={2}{12}];
\draw[dotted] (n011) -- (7,7.5);
\draw[dotted] (n011) -- (7,6.5);
\draw[dotted] (n014) -- (7,-6.5);
\draw[dotted] (n014) -- (7,-7.5);
\draw[dotted] (n025) -- (14, 7.5);
\draw[dotted] (n025) -- (14, 6.5);
\draw[dashed] (n026) -- node[above] {\footnotesize $M^1$} (16,6) [c0={3}{11}];
\draw[dashed] (n026) -- node[below] {\footnotesize $M^2$} (16,4) [c0={3}{12}]; 
\draw[dotted] (n027) -- (14,3.5);
\draw[dotted] (n027) -- (14,2.5);
\draw[dotted] (n028) -- (14, 1.5);
\draw[dotted] (n028) -- (14, 0.5);
\draw[dotted] (n029) -- (14, -0.5);
\draw[dotted] (n029) -- (14, -1.5);
\draw[dashed] (n0210) -- node[above] {\footnotesize $M^1$} (16,-2) [c0={3}{19}];
\draw[dashed] (n0210) -- node[below] {\footnotesize $M^2$} (16,-4) [c0={3}{20}]; 
\draw[dotted] (n0211) -- (14, -4.5);
\draw[dotted] (n0211) -- (14, -5.5);
\draw[dotted] (n0212) -- (14, -6.5);
\draw[dotted] (n0212) -- (14, -7.5);

\draw[dashed] (n0311) -- (19,6) [c0={N}{11}];
\draw[dashed] (n0312) -- (19,4) [c0={N}{12}];
\draw[dashed] (n0319) -- (19,-2) [c0={N}{19}];
\draw[dashed] (n0320) -- (19,-4) [c0={N}{20}];

 \draw (0,-9.5) -- node[above] {\footnotesize $u_0^1$}(6,-9.5) -- node[above] {\footnotesize $u_1^1,\ u_1^2,\ u_1^3,\ u_1^4$}(13,-9.5) -- node[above] {\footnotesize $\{ u_2^{j_2} \}_{j_2 = 1}^{16}$}(16, -9.5) -- (19,-9.5);    
  \foreach \y in {0,6,13,16,19}{
     \draw (\y,-9.6) -- (\y,-9.4);
     }
\draw (0,9) -- node[above] {\footnotesize $\mathcal{I}_0^1$}(6,9) -- node[above] {\footnotesize $\mathcal{I}_1^1,\mathcal{I}_1^2,\mathcal{I}_1^3,\mathcal{I}_1^4$}(13,9) -- node[above] {\footnotesize $\{\mathcal{I}_2 ^{j_2}\}_{j_2 = 1}^{16}$}(19,9);    
  \foreach \y in {0,6,13,19}{
     \draw (\y,8.9) -- (\y,9.1);
     }     
\draw [decorate,decoration={brace,amplitude=4pt,mirror},yshift = -5]
      (0,-9.7) -- (13,-9.7) node [midway,yshift=-0.3cm] {\footnotesize $Dual \; part$};
\draw [decorate,decoration={brace,amplitude=4pt,mirror},yshift = -5]
      (13,-9.7) -- (19,-9.7) node [midway,yshift=-0.3cm] {\footnotesize $Exploitation \; part$};                               
\end{tikzpicture}
\caption{Scenario tree for $L=2$: The dual part starts at a given initial condition and evolves, under a given input sequence, according to the $n_m = 2$ operating modes $M_1,M_2$, and the $N_s=2$ noise $w$ and parameter $\gamma$ samples. The exploitation part further predicts the trajectory using the current parameter and mode distribution for steps $k=2,\dots,N-1$.}
\label{fig:Figure1}
\end{figure}
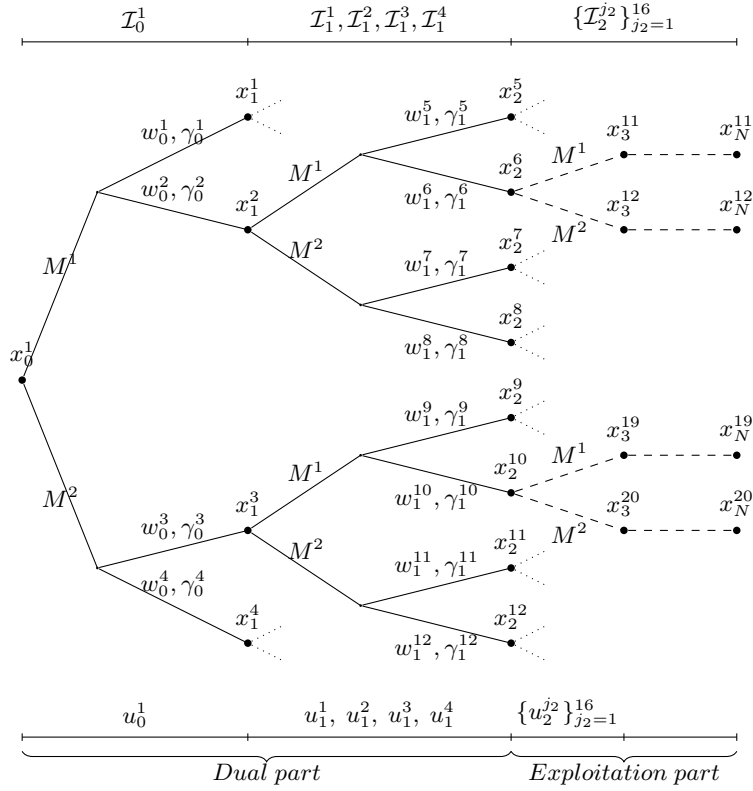

The dual part approximates~\eqref{eq:DP_recursion} by computing the expected values as averages over samples of the noise $w$ and of the parameter $\gamma$, weighted by the probability of each mode $M$. This induces an associated scenario tree depicted in Figure~\ref{fig:Figure1}, in which each state node $x_k^{j_k}$, denoted by a black circle, generates together with its corresponding input $u_k^{j_k}$ an associated subtree, where $j_k = 1, \dots, (N_sn_m)^k, \ k=0,\dots,L$. In each node, at every time step, there are two stages of branching. The first considers each potential mode $M$, while in the second, $N_s$ noise and parameter samples are realized, generating in total $N_s \times n_m$ leaf nodes. This procedure is repeated iteratively along a horizon of length $L$, resulting in the following update for each state sample 
\begin{equation}
x_{k+1}^{j_{k+1}} = \Phi^M(x_k^{P(j_{k+1})}, u_k^{P(j_{k+1})})\gamma_{k}^{j_{k+1}} + w_{k}^{j_{k+1}}, \quad j_{k+1} =  1, \dots, (N_sn_m)^{k+1} 
\label{eq:tree}
\end{equation}
where the index $j_{k+1}$ refers to the samples at stage $k+1$, and $P(\cdot)$ indicates the parent function.
The sampled state realizations obtained from~\eqref{eq:tree} are used to  construct the information vector $\mathcal{I}_k^{j_k}$ associated with each node. The dual effect is obtained by using~\eqref{eq:Bayesian_update} to compute both the distribution of $\gamma$, i.e.\ $p(\gamma^{M} \, | \, \mathcal{I}^{j_k}_{k}, M)$, from which we generate the parameter samples, as well as the probability of each mode $p(M \, | \, \mathcal{I}^{j_k}_{k})$, which is used to weigh the cost when averaging over all nodes of a time step. The expected values in~\eqref{eq:DP_recursion} are therefore substituted with weighted averages over these samples, which allows for unnesting the minimizations arising in the DP recursion, and optimizing over all the control input sequences in the tree at the same time, as was first proposed for handling parametric uncertainty in~\cite{EA19}. Therefore, while the DP recursion is not explicitly carried out, the input associated with each subtree is optimized while simultaneously considering future observations, and reacting to the corresponding sampled disturbance realizations, hence providing dual control. The optimization problem for the dual part is given by
\begin{equation}
   \tilde J_0(I_0^1) := \min_{\boldsymbol{u}_0, \dots, \boldsymbol{u}_{L-1}} \ \bar p_{j_0} l_0(x_0^1, u_0^1) + \frac{1}{N_s}\sum_{j_1=1}^{N_sn_m}  \bar p_{j_1} l_1(x_1^{j_1},u_1^{j_1}) + \dots  +  \frac{1}{N_s^{L}} \sum_{j_{L}=1}^{(N_sn_m)^{L}} \bar p_{j_{L}} \tilde J_L(\mathcal{I}_L^{j_L})  ,
  \label{eq:dual_problem}
  \end{equation}
where $\boldsymbol{u}_k = \{ u_k^1, \dots, u_k^{(N_sn_m)^k} \}$ refers to the collection of inputs for each stage of the problem, and $ u_k^{j_k} \in \mathcal{U}$ . The weight $\bar{p}_{j_k}$, assigned to each node, can be recursively obtained using the mode probability
\begin{equation}
\bar{p}_{j_{k+1}} = p(M|I_{k}^{P(j_{k+1})})\bar{p}_{P(j_{k+1})}, \quad k=0,\dots,L-1
\label{cost function weight} 
\end{equation}
with $\bar{p}_{j_0} = 1$. Details about the derivation of~\eqref{eq:dual_problem} can be found in \version{Appendix~\ref{app:dual part cost function}}{Appendix B in~\cite{EA19b}}.

\subsection{Exploitation Part}
\label{Exploitation Part}

In the exploitation part, we formulate the approximate cost-to-go $\tilde J_L(\mathcal{I}^{j_L}_L)$ for each scenario, by fixing the information collected until step $L$. Therefore, the current probability of each operating mode and its associated parameter distribution remain constant along each scenario in the exploitation part. Optimizing over these branches corresponds to solving a non-dual stochastic MPC problem
\begin{equation}
\begin{split}
& \tilde J_L(\mathcal{I}_L^{j_L}) = \min_{u_L^{j_L}, \dots, u_{N-1}^{j_L}}  \tilde J_L(\mathcal{I}_L^{j_L},u_{L:N-1}^{j_L}) , \\
& \tilde J_L(\mathcal{I}_L^{j_L},u_{L:N-1}^{j_L}) = \sum_{m=1}^{n_m}p(M^m|\mathcal{I}^{P(j_{L+1})}_L)\ \left( \mathbb{E}_{\gamma^M|M, w^M_L, \dots, w^M_{N-1}} \left[ l_L(x_L^{P(j_{L+1})}, u_L^{P(j_{L+1})}) + \right . \right. \\
&   + \sum_{k=L+1}^{N-1} l_k(x_k^{j_{L+1}}, u_k^{j_{L+1}}) + l_N(x_N^{j_{L+1}}) \ | \ \mathcal{I}_L^{P(j_{L+1})} \Big] \Big) 
\label{eq:exploitation_part}
\end{split}
\end{equation}
where the expected value with respect to the current mode probability, given the information at parent node $P(j_{L+1})$ with $j_{L+1} = 1,\dots, n_m(N_sn_m)^L$, is formulated explicitly. When the cost function is chosen to be either quadratic or linear, there exist analytic reformulations of the expected value in~\eqref{eq:exploitation_part} with respect to the first two moments of the predicted state. A common approach consists in considering only mean information, and is generally referred to as certainty equivalence control, for which we consider~\eqref{eq:model} evaluated at the state and parameter mean
\begin{subequations}\label{eq:certainty_equivalence}
\begin{align}
& \mu^{j_{L+1}}_{x_{L+1}}  = \Phi^M ( x^{P(j_{L+1})}_L , u_L^{P(j_{L+1})} ) \mu^{P(j_{L+1})}_{\gamma_L}
\label{eq:L_step} \\
& \mu^{j_{L+1}}_{x_{k+1}}  = \Phi^M ( \mu^{j_{L+1}}_{x_{k}}, u_k^{j_{L+1}} ) \mu^{P(j_{L+1})}_{\gamma_L}, \quad k=L+1, \dots, N-1  
\label{eq:all_steps}
\end{align}
\end{subequations}
where $\mu^{P(j_{L+1})}_{\gamma_L}$ is fixed at the value obtained at the last dual step $L$, and for parent node $P(j_{L+1})$.  A derivation of the model propagation in terms of state mean and variance information, using a first-order Taylor approximation of the dynamics around the state mean, is provided in \version{Appendix~\ref{app:exploitation part taylor}}{Appendix C in~\cite{EA19b}}.

\subsection{Final Dual MPC Problem}

The overall approximate dual control formulation provides a setup in which the two subproblems, namely the dual and exploitation part, can be merged into one optimization problem that optimizes over control sequences along the whole prediction horizon of length $N$, and which can be solved in a receding horizon fashion:
\begin{mini}[2]{\boldsymbol{u}_0,\dots,\boldsymbol{u}_{N-1}}{ \! \! \! \! \sum_{k=0}^{L-1} \frac{1}{(N_s)^k}\sum_{j_k=1}^{(N_sn_m)^k} \bar p_{j_k} l_k(x_k^{j_k},u_k^{j_k}) + \frac{1}{(N_s)^L}\sum_{j_L=1}^{(N_sn_m)^L} \bar p_{j_L} \tilde J_L(\mathcal{I}_L^{j_L},u_{L:N-1}^{j_L})}{\label{eq:overall_problem}}{}
\addConstraint{\! \! \! \! \eqref{eq:tree} }{}{\quad k=0,\dots,L-1}
\addConstraint{\! \! \! \! \gamma_k^{j_k}\text{ drawn from }\text{P}(\gamma^M | M, \mathcal{I}_{k}^{j_{k}}) }{}{\quad k=0,\dots,L-1}
\addConstraint{\! \! \! \! \eqref{cost function weight} }{}{\quad k=0,\dots,L-1}
\addConstraint{\! \! \! \!  \eqref{eq:certainty_equivalence} }{}{\quad k=L, \dots, N-1}
\addConstraint{\! \! \! \! u_k^{j_k} \in \mathcal{U} }{}{\quad k=0, \dots, N-1}
\end{mini}
Parameter samples $\gamma^{j_k}_k$ are generated online as affine transformations of samples from a standard normal distribution, using exact mean and covariance provided by~\eqref{eq:Bayesian_update_param}, and can be drawn before optimization, such that standard solvers for gradient-based optimization can be used. 
\begin{remark}
The formulation of problem~\eqref{eq:overall_problem} offers the possibility to use deferred dual steps, i.e. inserting dual steps after a number of exploitation steps. This is useful for instance when the system requires several time steps to achieve good excitation, such that identification later in the prediction can be significantly more informative.
\end{remark}

\section{Simulation Example}

The following example demonstrates the developed dual MPC (DMPC) for the problem of controlling an aircraft to a desired altitude, with a reference change of $50 \ m$. The nominal operating mode of the system follows the continuous-time linearized longitudinal dynamics of the Cessna Citation aircraft in~\citep{JMM02}. 
The state of the system $x_k$ describes the evolution of the angle of attack, the pitch angle, the pitch rate and the altitude. The input $u_k$ corresponds to the elevator angle, and is constrained to lie between $\pm 0.2 \ rad$. As a simulation scenario, we consider the case in which a fault occurs to the actuator some time before the reference change, leading to a decrease in gain of $75 \%$. We identify two operating modes of the system with $M^1$ being nominal and $M^2$ being the fault mode, each with a corresponding uncertain parameter $\gamma^M$ describing the actuator gain
\begin{equation*}
 M : \ x_{k+1} = Ax_k + \gamma^MBu_k + w_k, \quad M \in \{ M^1, M^2\}
\end{equation*}
where $A,B$ are the discretized system matrices using a sampling time $T_s =0.2\ s$. We assume that $\gamma^2$ is subject to larger uncertainty than $\gamma^1$, as it is hard to know a priori how much loss in gain has occurred. For this reason we choose the prior distribution for each parameter as $\gamma^1 \sim \mathcal{N}(0.95,(0.03)^2)$ and $\gamma^2 \sim \mathcal{N}(0.4,(0.1)^2)$. Both modes are subject to additive process noise $w_k \sim \mathcal{N}(0,(0.3)^2) $. Compared with the use of a unimodal parameter distribution with a single mode, explicitly considering a failure case provides faster model identification, as shown for instance in~\citep{JDB99}.
\begin{figure}[h]
  \centering
    \input{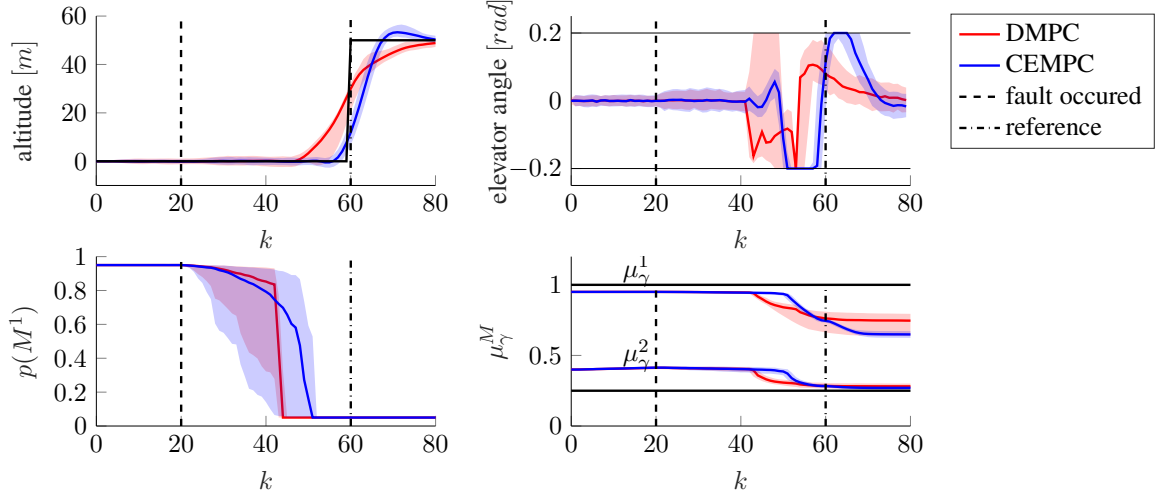}
\caption{Altitude reference change. Each plot depicts the DMPC (in red) and CEMPC (in blue) performance averaged over 200 noise realizations, for which the median is highlighted in bold. The shaded area around the median is bounded by the 5 and 95 percent quantiles. \emph{Top left}: altitude state trajectory. \emph{Top right}: elevator input. \emph{Bottom left}: nominal mode probability. \emph{Bottom right}: parameter mean for both modes.}
\label{fg:figure}
\end{figure}
We generate a DMPC with dual lookahead $L=1$, $n_m = 2$ operating modes, and $N_s = 2$ noise and parameter realizations, which result in $N_s \times n_m =4$ scenarios, and the overall prediction horizon is $N=20$. We compare the DMPC with a non-dual, passive adaptive controller, referred to as CEMPC. This is based on a certainty equivalence approach, i.e. a deterministic receding horizon controller that computes open-loop sequences, and in which the probability of the operating mode and the underlying distribution of the associated unknown parameter are updated in closed-loop following~\eqref{eq:Bayesian_update}. Figure~\ref{fg:figure} shows the performance of both controllers for the specified task, where the first dashed line refers to the fault occurring at time step $k=20$, and the second dashed-dotted line corresponds to the reference change at time step $k=60$. For DMPC, the elevator input (top right) explores aggressively as soon as the reference change enters the prediction horizon at time step $k=40$, and displays goal-oriented control, i.e. fast identification of a fault ensuring adequate performance by initiating the pitching maneuver early. On the other hand, the non-dual CEMPC input does not excite the system sufficiently in advance, causing input saturation and late response to the reference change. The probability of the nominal mode (bottom left) in the DMPC shifts in one time step at $k=40$ to $p(M^1) = 0.05$, which corresponds to the chosen lower bound (cf. Remark~\ref{capping remark}). The DMPC bimodal distribution defined by the two parameters $\gamma^1$ and $\gamma^2$ starts to adapt due to introduced excitement at $k = 40$ towards the true parameter $\gamma^2 = 0.25$, associated with the fault mode active at the reference change. Convergence of $\mu^1_{\gamma}$ slows down as the aircraft approaches the reference since input excitation decreases, i.e. further identification is no longer needed. Similarly, the estimate for $\gamma^2$ stagnates due to little excitation, but would eventually converge to $0.25$. The DMPC controller therefore provides rapid identification of the fault condition, and of the specific loss of the actuator gain. 
\acks{This work was supported by the Swiss National Science Foundation under grant no. PP00P2 157601 / 1 and by the CTI-Project Nr. 25959.2 PFIW-IW.}

\bibliography{l4dc2020-dualcontrol}

\version{
  
   \appendix

   \section{Parameter Update - Bayesian Estimation}\label{app:bayesian estimation}

  Assume that the distribution of parameter $\gamma^M \in \mathbb{R}^{n_\gamma^M}$ at time step $k$ is given by
  \begin{equation*}
  p(\gamma^M | \mathcal{I}_k , M) =  \mathcal{N}( \ \mu_{\gamma,k}^M \ ,  \ \Sigma^M_{\gamma,k} \ ).
  \end{equation*}
  Since the additive process noise in~\eqref{eq:model} is Gaussian distributed with $w^M_k \sim \mathcal{N}(0,\Sigma^M_w)$, the likelihood of the state at time step $k+1$ can be computed as
  \begin{equation*}
  p(x_{k+1} | u_k, \mathcal{I}_k, M, \gamma^M) = \mathcal{N}( \ \Phi^M(x_k,u_k)\gamma^M, \ \Sigma^M_w \ ).
  \end{equation*}
  The update of the posterior distribution $p(\gamma^M | \mathcal{I}_{k+1} , M)$ in~\eqref{eq:Bayesian_update_param}, given the information vector at time step $k+1$, can therefore be evaluated in closed form since the Gaussian distribution is self-conjugate. The updated parameter mean and variance result in~\citep{CB06}:
  \begin{equation*}
  \begin{split}
  & \left[ \Sigma_{\gamma,k+1}^M \right]^{-1} = \left[ \Sigma_{\gamma,k}^M \right]^{-1} + \Phi^M(x_k, u_k)^T \ \left[ \Sigma_w^{M} \right]^{-1} \ \Phi^M(x_k, u_k), \\
  & \mu_{\gamma,k+1}^M = \Sigma_{\gamma,k+1}^M \left( \left[ \Sigma_{\gamma,k}^M \right]^{-1} \mu_{\gamma,k}^M + \Phi^M(x_k,u_k)^T \ \left[ \Sigma_w^{M}\right]^{-1} \ x_{k+1}\right).
  \end{split}
  \end{equation*}
  The update of the mode probability in~\eqref{eq:Bayesian_update_model} is enabled by using the evidence (or marginal likelihood) in~\eqref{eq:Bayesian_update_param}, which can be computed by marginalizing with respect to $\gamma^M$:
  \begin{equation*}
  \begin{split}
  p(x_{k+1} | u_k, \mathcal{I}_k, M) & = \int{ p(x_{k+1} | u_k, \mathcal{I}_k, M, \gamma^M) \  p(\gamma^M | \mathcal{I}_k , M) \ d\gamma^M} \\
  & = \mathcal{N}( \ \Phi^M(x_k,u_k) \mu_{\gamma,k}^M \ , \ \Sigma^M_w \ + \ \Phi^M(x_k,u_k) \ \Sigma^M_{\gamma,k} \Phi^M(x_k,u_k)^T ),
  \end{split}
  \end{equation*}
  which can be computed in closed form since the convolution of two Gaussian distributions is Gaussian itself. The updated information $\mathcal{I}_{k+1}$ is then used to update the probability of mode $M$ as:
  \begin{equation*}
  p(M|\mathcal{I}_{k+1}) = \frac{\mathcal{N}( \ \Phi^M(x_k,u_k) \mu_{\gamma,k}^M \ , \ \Sigma^M_w \ + \ \Phi^M(x_k,u_k) \  \Sigma^M_{\gamma,k} \Phi^M(x_k,u_k)^T ) \ p(M|\mathcal{I}_{k})}{\sum_{m=1}^{n_m}{ \ \mathcal{N}( \ \Phi^{M^m}(x_k,u_k) \mu_{\gamma,k}^{M^m} \ , \ \Sigma^{M^m}_w \ + \ \Phi^{M^m}(x_k,u_k) \  \Sigma^{M^m}_{\gamma,k} \Phi^{M^m}(x_k,u_k)^T ) \ p(M^{m}|\mathcal{I}_{k}) } },
  \end{equation*}
  where $n_m$ is the total number of modes.

   \section{Dual Part - Cost Function}\label{app:dual part cost function}

   The formulation of the dual part in~\eqref{eq:dual_problem} is based on the scenario tree in Figure~\ref{fig:Figure1}, which is used to obtain an approximation of the DP recursion in~\eqref{eq:DP_recursion}. We outline the procedure for obtaining~\eqref{eq:dual_problem} when the length of the dual horizon is $L=2$. Assume the cost-to-go at time step $k=1$ is available, i.e. $\tilde J_1(\mathcal{I}_1)$. The cost-to-go at time step $k=0$ can be expressed as:
   \begin{equation*}
   \tilde J_0(\mathcal{I}_0) = \min_{u_0} \ l(x_0, u_0) + \mathbb{E}_{M,\gamma^M,w^M_0} \left[ \tilde J_1(\mathcal{I}_1) \ | \ \mathcal{I}_0 \right].
   \end{equation*}
   Using the law of iterated expectations, and explicitly formulating the expected value with respect to the mode, we obtain the following expression
   \begin{equation*}
   \tilde J_0(\mathcal{I}_0) = \min_{u_0} \ l(x_0, u_0) + \sum_{m=1}^{n_m} \ p(M^m | \mathcal{I}_0) \ \left( \mathbb{E}_{\gamma^{M^m}|M^m,w^{M^m}_0} \left[ \tilde J_1(\mathcal{I}_1) \ | \ \mathcal{I}_0 \right] \right),
   \end{equation*}
   where $n_m$ is the total number of modes. The expected value with respect to parameter $\gamma^M$, given mode $M$, and noise $w_0^M$ is evaluated by averaging over $N_s$ samples of $\gamma^M \sim \mathcal{N}(\mu_{\gamma,0}^M, \Sigma_{\gamma,0}^M)$ and $w^M \sim \mathcal{N} (0,\Sigma^M_0)$:
   \begin{equation*}
   \tilde J_0(\mathcal{I}_0) = \min_{u_0} \ l(x_0, u_0) + \sum_{m=1}^{n_m} \ p(M^m | \mathcal{I}_0) \ \left( \frac{1}{N_s} \sum_{l=1}^{N_s} \tilde J_1(\mathcal{I}^l_1) \ \right),
   \end{equation*}
   where the index $l$ iterates over the $N_s$ nodes associated with each mode $M$. It is possible to express the nested sum over the $n_m$ modes and the $N_s$ samples as one sum, by using a global index $j_1 = 1, \dots, N_sn_m$ that iterates over all nodes at $k=1$. Therefore, we define a weight $\bar p_{j_1}$ that is equal to $p(M^m|\mathcal{I}_0)$, where mode $M^m$ generates the corresponding node $j_1$, and use it to express the sum indexed by $j_1$:
   \begin{equation*}
   \tilde J_0(\mathcal{I}_0) = \min_{u_0} \ l(x_0, u_0) + \frac{1}{N_s} \sum_{j_1=1}^{N_sn_m} \ \bar p_{j_1}  \ \tilde J_1(\mathcal{I}^{j_1}_1).
   \end{equation*} 
   We repeat this procedure similarly for time step $k=1$, assuming that the cost-to-go at time step $k=2$ is available for each node $j_1$:
   \begin{equation*}
   \begin{split}
   \tilde J_1(\mathcal{I}^{j_1}_1 ) & = \min_{u_1^{j_1}} \ l(x_1^{j_1}, u_1^{j_1}) + \mathbb{E}_{M,\gamma^M,w^M_1} \left[ \tilde J_2(\mathcal{I}^{j_1}_2) \ | \ \mathcal{I}^{j_1}_1 \right] \\
   & =  \min_{u_1^{j_1}} \ l(x_1^{j_1}, u_1^{j_1}) + \sum_{m=1}^{n_m} \ p(M^m | \mathcal{I}^{j_1}_1) \ \left( \frac{1}{N_s} \sum_{l=1}^{N_s} \tilde J_2(\mathcal{I}^{j_1, l}_2) \  \right),
   \end{split}
   \end{equation*}
   where the index $l$ iterates over the $N_s$ nodes associated with each mode $M$, branching from node $j_1$. The cost $\tilde J_0(\mathcal{I}_0)$ is expressed as:
   \begin{equation*}
   \begin{split}
   & \tilde J_0(\mathcal{I}_0) = \\
   & \min_{u_0} \ l(x_0, u_0) + \frac{1}{N_s} \sum_{j_1=1}^{N_sn_m} \ \bar p_{j_1}  \ \left( \min_{u_1^{j_1}} \ l(x_1^{j_1}, u_1^{j_1}) + \sum_{m=1}^{n_m} \ p(M^m | \mathcal{I}^{j_1}_1) \ \left( \frac{1}{N_s} \sum_{l=1}^{N_s} \tilde J_2(\mathcal{I}^{j_1, l}_2) \  \right) \right).
   \end{split}
   \end{equation*}
   The nested sums over the $N_sn_m$ nodes at $j_1$, and the $M^m$ modes and $N_s$ samples for each $j_1$, can again be merged into one sum in which a global index $j_2 = 0, \dots, (N_sn_m)^2$ iterates over all the nodes at time step $k=2$. We define the weight $\bar p_{j_2} = p(M | \mathcal{I}^{P(j_{2})}_1) \bar p_{P(j_2)}$, where $P(j_2)$ denotes the node $j_1$ corresponding to the parent of $j_2$, and obtain:
   \begin{equation*}
   \tilde J_0(\mathcal{I}_0) = \min_{u_0} \ l(x_0, u_0) + \frac{1}{N_s} \sum_{j_1=1}^{N_sn_m} \  \min_{u_1^{j_1}} \ \bar p_{j_1}  \  l(x_1^{j_1}, u_1^{j_1}) + \frac{1}{N_s^2} \sum_{j_2=1}^{(N_sn_m)^2} \ \bar p_{j_2} \tilde J_2(\mathcal{I}^{j_2}_2) .
   \end{equation*}
   This procedure can be extended to the case of a general dual horizon of length $L$, which results in the formulation given in~\eqref{eq:dual_problem}.

   \section{Exploitation Part - First Order Taylor Approximation}\label{app:exploitation part taylor}

   A first-order Taylor approximation of~\eqref{eq:model} around the state and parameter mean is computed by extending the state with the (constant) parameter dynamics, where we omit the sample index for notational simplicity:
  \begin{equation*}
\begin{split}
& \begin{bmatrix}
x_{k+1} \\
\gamma_{k+1}
\end{bmatrix} \approx \\
& \begin{bmatrix}
\Phi^M( \mu_{x_k} , u_k) \mu_{\gamma_L}  \\
\mu_{\gamma_L}
\end{bmatrix} + 
\begin{bmatrix}
w_k \\
0
\end{bmatrix} + 
\begin{bmatrix}
\nabla_x [ \Phi^M( \mu_{x_k} , u_k) \mu_{\gamma_L}] &  \Phi^M( \mu_{x_k} , u_k) \\
0 & \mathbb{I}
\end{bmatrix}
\begin{bmatrix}
x_{k} \ - \  \mu_{x_k} \\
\gamma_k \ - \ \mu_{\gamma_L}
\end{bmatrix}, 
\end{split} 
\end{equation*}
for $k=L+1, \dots, N-1$, and where $\mu_{\gamma_L}$ is the mean at the last dual step $k=L$. This provides update equations for both mean and variance based on the properties of affine transformations of Gaussian distributed variables. At time step $k=L$ we have the following initialization:
 \begin{equation*}
\begin{split}
\begin{bmatrix}
\mu^{j_{L+1}}_{x_{L+1}} \\
\mu^{j_{L+1}}_{\gamma_{L+1}}
\end{bmatrix} & = \begin{bmatrix}
\Phi^M( x^{P(j_{L+1})}_L , u^{P(j_{L+1})}_L) \mu^{P(j_{L+1})}_{\gamma_L}  \\
\mu^{P(j_{L+1})}_{\gamma_L}
\end{bmatrix}, \\
\Sigma^{j_{L+1}}_{L+1} & = \bar \Sigma^{P(j_{L+1})}_w + \bar A_L \Sigma^{P(j_{L+1})}_{L} \bar A_L^T , 
\end{split}  
\end{equation*}
where 
 \begin{equation*}
\begin{split}
\Sigma^{P(j_{L+1})}_L & = \begin{bmatrix}
0  & 0 \\
0 & \Sigma^{P(j_{L+1})}_{\gamma_L}
\end{bmatrix}, \\
\bar A_L & = \begin{bmatrix}
 \nabla_x [ \Phi^M( x^{P(j_{L+1})}_L , u^{P(j_{L+1})}_L) \mu^{P(j_{L+1})}_{\gamma_L}] &  \Phi^M( x^{P(j_{L+1})}_L , u^{P(j_{L+1})}_L) \\
0 & \mathbb{I}
\end{bmatrix}, \\
\bar \Sigma^{P(j_{L+1})}_w & = \begin{bmatrix}
\Sigma^{P(j_{L+1})}_w & 0 \\
0 & 0
\end{bmatrix},
\end{split}  
\end{equation*}
where the index at step $k=L+1$  is $j_{L+1} = 1, \dots, n_m(N_sn_m)^L$, and $P(j_{L+1})$ denotes the node $j_L$ corresponding to the parent of $j_{L+1}$. For the subsequent time steps $k= L+1, \dots, N-1$, we have:
\begin{equation*}
\begin{split}
\begin{bmatrix}
\mu^{j_{L+1}}_{x_{k+1}} \\
\mu^{j_{L+1}}_{\gamma_{k+1}}
\end{bmatrix} & = \begin{bmatrix}
\Phi^M( \mu^{j_{L+1}}_{x_k} , u^{j_{L+1}}_k) \mu^{j_{L+1}}_{\gamma_k} \\
\mu^{j_{L+1}}_{\gamma_k}
\end{bmatrix} , \\
\Sigma^{j_{L+1}}_{k+1} & = \bar \Sigma^{j_{L+1}}_w + \bar A_k \Sigma^{j_{L+1}}_{k} \bar A_k^T , \end{split}  
\end{equation*}
where
\begin{equation*}
\begin{split}
\Sigma^{j_{L+1}}_k & = \begin{bmatrix}
\Sigma^{j_{L+1}}_{x_k}  & \Sigma^{j_{L+1}}_{x_k, \gamma_k} \\
\Sigma^{j_{L+1}}_{\gamma_k, x_k} & \Sigma^{j_{L+1}}_{\gamma_k}
\end{bmatrix}, \\
\bar A_k & = \begin{bmatrix}
 \nabla_x [ \Phi^M( \mu^{j_{L+1}}_{x_k} , u^{j_{L+1}}_k) \mu^{j_{L+1}}_{\gamma_k}] &  \Phi^M( \mu^{j_{L+1}}_{x_k} , u^{j_{L+1}}_k) \\
0 & \mathbb{I}
\end{bmatrix}, \\
\bar \Sigma^{j_{L+1}}_w & = \begin{bmatrix}
\Sigma^{j_{L+1}}_w & 0 \\
0 & 0
\end{bmatrix}.
\end{split}
\end{equation*}

}

\end{document}